\documentclass[         %
aps,                    
prd,                    
showpacs,               
preprint,              
nofootinbib,            
showkeys,               %
preprintnumbers,        %
floatfix]               
{revtex4}               
\usepackage{graphicx,longtable} 

\begin{document}
\title{Constraints on Neutrino Parameters from Neutral-Current 
Solar Neutrino Measurements}
\author{        A.~B. Balantekin}
\email{         baha@nucth.physics.wisc.edu}
\author{        H. Y\"{u}ksel}
\email{         yuksel@nucth.physics.wisc.edu}
\affiliation{  Department of Physics, University of Wisconsin\\
               Madison, Wisconsin 53706 USA }
\date{\today}
\begin{abstract} 
We generalize the pull approach to define the $\chi^2$ function 
to the analysis of the data with correlated statistical errors. 
We apply this method to the analysis of the Sudbury Neutrino 
Collaboration data obtained in the salt-phase. 
In the global analysis of all the solar neutrino and KamLAND data 
we find the best fit (minimum $\chi^2$)  
values of neutrino parameters to be $\tan^2 \theta_{12} \sim 0.42$ 
and $\delta m_{12}^2 \sim 7.1 \times 10^{-5}$ eV$^2$.
We confirm that the maximal mixing is 
strongly disfavored while the bounds on $\delta m_{12}^2$ 
are significantly strengthened. 
\end{abstract}
\medskip
\preprint{} 
\maketitle

As the solar neutrino physics moves from the discovery stage to the 
precision measurements stage increasingly more data are becoming 
available for a critical analysis. Several experimental groups 
recently announced new results. The Sudbury Neutrino 
Observatory (SNO) 
Collaboration announced results from the 
neutral-current measurements 
in the salt phase of the experiment \cite{yenisno}. 
A new, very precise measurement 
of the astrophysical S-factor for the reaction 
$^7$Be(p,$\gamma$)$^8$B was performed by the Seattle-TRIUMF group 
\cite{Junghans:2003bd}, updating their previous result 
\cite{Junghans:2001ee}. 

Earlier SNO charged-current (CC) measurements 
\cite{Ahmad:2001an,Ahmad:2002ka} 
confirmed the deficiency in the solar neutrino flux and 
helped narrow the neutrino parameter space. 
Earlier SNO
neutral-current (NC)
measurements \cite{Ahmad:2002jz} yielded a total (all flavors) $^8$B 
solar neutrino flux which is in very good agreement with 
the Standard 
Solar Model (SSM) predictions \cite{Bahcall:2000nu}. In a parallel 
development the reactor antineutrino disappearance observed by the 
KamLAND Collaboration \cite{Eguchi:2002dm} found to favor the same 
region of the neutrino parameter space (the so-called Large 
Mixing Angle 
(LMA) region) as the solar neutrino experiments. In a 
sense KamLAND and 
SNO are complementary experiments with KamLAND being more sensitive 
to the difference between squares of the neutrino masses,
$\delta m^2$, and SNO providing increasingly more precise constraints 
on the mixing angle as well as $\delta m^2$. 

The new SNO NC measurement is able to determine the total 
active solar 
neutrino flux in a model-independent way. 
In particular no assumptions need to be made about the 
energy dependence 
of the flux. This feature makes the new data very valuable 
in restricting 
SSM or neutrino properties.

In a previous paper \cite{Balantekin:2003dc} we presented a 
global analysis 
of all the solar neutrino and KamLAND data using a 
covariance approach. 
The aim of this paper is to perform a similar analysis using 
the pull 
approach. We use the pull 
approach in the form proposed and implemented by Fogli {\em et al} 
\cite{Fogli:2002pt}. Although the covariance and pull approaches are 
strictly equivalent, the pull approach seems to provide a 
treatment of statistical and systematic uncertainties which is more 
transparent, easier to implement, and computationally 
simpler. In our analysis we include correlations between statistical 
errors. 



We now enumerate the changes we incorporated in this paper 
as compared with our previous global analysis. 
The calculations of neutrino propagation in matter are done as 
described in 
Ref. \cite{Balantekin:2003dc}. Since the SNO 
Collaboration has not yet announced a live-time distribution 
for zenith
angles, we used the live-time information previously 
provided in the SNO data page \cite{HOWTOSNO} to calculate Earth 
regeneration effects and the 24-hour average neutrino survival 
probabilities.  
In our calculations we ignored experimental correlations 
between SNO's 
data sets as recommended by the SNO Collaboration \cite{HOWTOSNO2}. 
They emphasized that even though treatment of these 
correlations require 
extensive knowledge of the SNO detector, they have very 
little impact 
on determining confidence level regions. 

The data set used in this analysis includes
the average rate of the gallium experiments
(SAGE \cite{Abdurashitov:2002nt}, GALLEX \cite{Hampel:1998xg}, GNO 
\cite{Altmann:2000ft}), the  total rate of the Homestake chlorine
experiment \cite{Cleveland:nv}, 44 data points from the 1496 days
SK zenith-angle-spectrum \cite{Fukuda:2002pe}, 34 data points 
from the
2002 SNO day-night-spectrum \cite{Ahmad:2002ka} as well as the 
newly-released SNO NC, CC, and electron scattering (ES) 
data \cite{yenisno}.   
We calculated the CC cross section  
using the effective field approach \cite{Butler:1999sv}, with 
radiative corrections as described in Ref. 
\cite{Kurylov:2002vj}. In our 
calculations we used the counter-term value of $L_{1A} = 4.5$ fm$^3$ 
\cite{Chen:2002pv,Balantekin:2003ep}. 
For new salt phase SNO data we calculated the detector-averaged 
cross-sections using updated detector response function given in 
Ref. 
\cite{yenisno} and higher threshold of 5.5 MeV as described in 
the SNO data page \cite{HOWTOSNO2}. 
The NC, CC, and ES fluxes are added as new data points. 
The systematic correlations of these new measured NC, CC, and ES 
fluxes are
taken into account as described in \cite{HOWTOSNO2}. The main 
systematic 
sources of errors, like  
energy scale, 
radial accuracy, isotropy mean, radial energy bias, internal 
background 
neutrons, neutron capture, etc. are taken into account with the 
appropriate correlation coefficients given in \cite{HOWTOSNO2}. 

Following the suggestion of Ref. \cite{Bahcall:2002zh}, we added 
cross section errors first linearly and
then quadratically in the parts contributing to the
low and high energy parts of for radiochemical experiments. 
This change in  cross section treatment resulted in a slight 
reduction of the allowed region contours. 
SNO collaboration has provided a 
statistical correlation matrix for the salt phase. 
The analysis of Ref. \cite{Fogli:2002pt} proves that for 
uncorrelated statistical errors the pull and covariance approaches 
yield identical $\chi^2$ values. 
In the Appendix we write down the appropriate relations when  
statistical errors are correlated and 
prove that Fogli {\em et al.} result can be generalized   
to the correlated statistical errors.
In our calculations we used the SSM
systematics and fluxes 
\cite{Bahcall:2000nu}.

We calculated the theoretical CC flux using the prescription 
given in the SNO data 
page \cite{HOWTOSNO2}: 
\begin{equation}
\frac{\int_0^{\infty} \int_0^{\infty} \int_{5.5}^{\infty} \phi_{SSM} 
(E_{\nu}) P_{ee} \frac{d \sigma}{dT_e} (E_{\nu},T_e) R(T_e,T) 
dE_{\nu} dT_e 
dT}{\int_0^{\infty} \int_0^{\infty} \int_{5.5}^{\infty} \phi_{SSM} 
(E_{\nu}) \frac{d \sigma}{dT_e} (E_{\nu},T_e) R(T_e,T) dE_{\nu} dT_e 
dT},  
\label{1}
\end{equation}
where $\phi_{SSM}$ is the SSM flux, $P_{ee}$ is the electron 
neutrino 
survival probability, and $R$ is the energy response function.  
This result was used to calculate the CC/NC double ratio (see e.g. 
Ref. \cite{Maris:2001tg}). 

The collective effect of the fractional systematic errors of the SSM 
input 
parameters appear as a shift of the neutrino fluxes:
\begin{equation}
\Phi_i \rightarrow f_i \Phi_i,
\label{3}
\end{equation}
where the index $i$  represents the neutrino source (pp, $^8$B, 
etc.). 
The shift $f_i$ can easily be calculated in the pull method (cf. 
Eq. (16) of Ref. \cite{Fogli:2002pt}). In the covariance approach 
one can allow the 
$^8$B flux to float freely. The shift in the pull approach   
provides a similar, but not identical function.

All graphs are shown with several 
confidence levels.
In the figures shown the darkest shaded areas are the 90 
\% confidence 
level regions. Lighter shaded areas progressively include the 
95 \%, 99 \%,  and
99.73 \% confidence level regions. In this manner, for example 
the entire shaded region in a given graph is 
the 99.73 \% confidence level region. 
Best fit points are marked by a dagger.
Isolines are clearly marked with corresponding values of the 
quantities which are being examined.


In order to compare the results before and after adding the SNO 
salt phase 
data, in Fig. {\ref{fig:1} we show the LMA allowed region of solar 
neutrino parameter space.  In this figure the CC/NC ratio 
isocountours, which are calculated for SNO experiment
 2002 data  with $T_{e,th}=5$ MeV, are also shown. 
 At best fit (marked by a cross), the value of this  ratio is $0.35$.

In Fig. \ref{fig:2} we show the allowed confidence levels in the 
neutrino parameter space using only the total NC, CC, and ES fluxes 
measured in the SNO salt phase measurement. This graph illustrates the 
crucial role of the additional spectrum information in constraining the 
neutrino parameters even after the neutral current measurements fix 
the $^8$B flux.

In Fig. \ref{fig:3} we show the allowed confidence levels in the LMA 
region of neutrino parameter 
space when chlorine, average rate of gallium experiments, SK 
zenith-spectrum
and  SNO salt phase measurement of NC, CC and ES data are used. 
In the calculations leading to this graph we excluded the 
SNO day-night spectrum data from 2002 in  order to
better understand the impact of new salt phase measurement. 
This figure can be compared to the Fig. \ref{fig:1}. 
Fig. \ref{fig:3} uses the same input information as Fig. \ref{fig:1} except 
the SNO day-night information is replaced by the
salt-phase results. Thus the relative impacts of the two data sets 
(salt-phase 
vs. day-night spectrum) in constraining the allowed neutrino parameter space 
are exhibited.  
Already the LOW 
solution is excluded at the 99.78 \% confidence level. Clearly 
the information 
from the SNO salt phase helps significantly shrink the allowed 
region. 

The global analysis of all available solar neutrino data is 
illustrated in 
Fig. \ref{fig:4}. (This figure is similar to Fig. \ref{fig:3} 
except that 
the SNO day-night spectrum data are also included).  
We find the best fit (minimum $\chi^2$)  
values of neutrino parameters to be $\tan^2 \theta_{12} \sim 0.42$ 
and $\delta m_{12}^2 \sim 6.7 \times 10^{-5}$ eV$^2$ 
from our combined analysis of all the solar neutrino data. 
Our minimum $\chi^2$ value is 70.5 for 83 data points and 2 
parameters. The contours of the CC/NC ratio, which provides 
information about the 
electron neutrino survival probability, are also shown. Our best fit 
value 
for this ratio is $0.33$. This is in good agreement with the SNO 
value of 
$0.306 \pm 0.026 ({\rm stat}) \pm 0.024 ({\rm syst})$. 

The global analysis of all available solar neutrino plus the KamLAND 
data are  
shown in Figs. \ref{fig:6} and \ref{fig:7}. In Fig. \ref{fig:6} the 
$^8$B 
flux is fixed at the SSM value whereas in Fig. \ref{fig:7} it is 
taken as a 
free parameter. In Figure \ref{fig:6} we also show 
the ratio of this shifted $^8$B flux to the SSM value.
The best fit of the shifted flux corresponds to 1.02 times the SSM 
value. In both graphs 
we find the best fit (minimum $\chi^2$)  
values of neutrino parameters to be $\tan^2 \theta_{12} \sim 0.42$ 
and $\delta m_{12}^2 \sim 7.1 \times 10^{-5}$ eV$^2$. Even though 
the allowed 
regions of the neutrino parameter space are quite similar in shape, 
allowing 
the $^8$B flux float freely slightly enlarges the allowed region as 
one would expect.

In conclusion one observes that 
the new SNO NC, CC, and ES data fluxes are statistically correlated 
since they are derived from a single fit to the data.
In this paper we applied the pull method generalized to incorporate  
correlated statistical 
experimental errors to the analysis of the SNO data. We confirm that 
the neutrino parameters are much better constrained after the 
addition 
of the SNO salt-phase data. In particular the maximal mixing is 
strongly disfavored while the bounds on $\delta m_{12}^2$ are  
significantly strengthened. 

\section*{ACKNOWLEDGMENTS}
We thank Y.D. Chan, K.T. Lesko, A.W.P. Poon, R.G.H. Robertson, 
and J. 
Wilkerson for many discussions. One of us (H.Y.) would like thank  
INPA Neutrino Astrophysics Group for their hospitality.  
This work was supported in part by the U.S. National Science
Foundation Grant No.\ PHY-0244384 and in part by
the University of Wisconsin Research Committee with funds granted by
the Wisconsin Alumni Research Foundation. 

\section*{Pull Approach with Correlated Statistical Errors}

Folllowing the notation of 
Fogli {\em et al.} \cite{Fogli:2002pt} 
we designate the statistical errors for the observable 
$R_n$ by $u_n$, and the systematic error for the source $k$ for 
this observable by $c_n^k$. They also define scaled quantities 
\begin{equation}
\label{2a}
\Delta_n = \frac{R_n^\mathrm{exp} - R_n^\mathrm{th}}{u_n} ,
\end{equation}
and
\begin{equation}
\label{2b}
q^k_n = \frac{c^k_n}{u_n} .
\end{equation}
Thus for the covariance approach one can write
\begin{equation}
\label{2c}
\chi^2_\mathrm{covar} = \sum_{n,m} \Delta_n
[\rho_{nm} + {\scriptstyle\sum_k} q_n^k q_m^k]^{-1} \Delta_m  ,
\end{equation}
where $\rho_{nm}$ is the statistical covariance matrix including
 the correlated 
errors. For the pull approach one has 
\begin{equation}
\label{2d}
\chi^2_\mathrm{pull} = \min_{\{\xi_k\}} \left[\sum_{n,m} 
(\Delta_n - {\scriptstyle\sum_k} q_n^k\, \xi_k) \rho^{-1}_{nm}
(\Delta_m - {\scriptstyle\sum_{k'}} q_m^{k'}\, \xi_{k'}) 
 + \sum_{k}
\xi^2_k\right] ,
\end{equation}
where $\xi_k$ is the univariate Gaussian random variable introduced in 
the pull approach. Minimizing Eq. (\ref{2d}) with respect to $\xi_k$ we 
obtain
\begin{equation}
\label{2e}
  \xi_k=\sum_{h} S_{kh} \sum_{n,m} \rho^{-1}_{nm} \Delta_m q^h_n ,
\end{equation}
where
\begin{equation}
\label{2f}
S_{kh} = [\delta_{kh} + {\scriptstyle 
\sum_{n,m}} q_n^k \rho^{-1}_{nm} q_m^h]^{-1}\ .
\end{equation}
Defining the matrices $B$ and $D$ through their matrix elements
\begin{equation}
\label{2g}
B_{nm} = \sum_h q_n^h q_m^h, 
\end{equation}
and
\begin{equation}
\label{2gg}
D_{nm} = \sum_{k,h} S_{kh} q_n^h q_m^h, 
\end{equation}
one can show that the covariance matrix is $\rho + B$. Using the 
definition given in Eq. (\ref{2gg}) it follows that the inverse of the 
covariance matrix is given by
\begin{equation}
\label{2h}
(\rho + B)^{-1} = \rho^{-1} - \rho^{-1} D \rho^{-1} . 
\end{equation}
Substituting Eqs. (\ref{2e}) and (\ref{2h}) into Eqs. (\ref{2c}) and 
(\ref{2d}) after some algebra one gets 
\begin{equation}
\label{2i}
\chi^2_\mathrm{covar} \equiv \chi^2_\mathrm{pull}  .
\end{equation}



\newpage

\begin{figure}
\includegraphics[scale=.7]{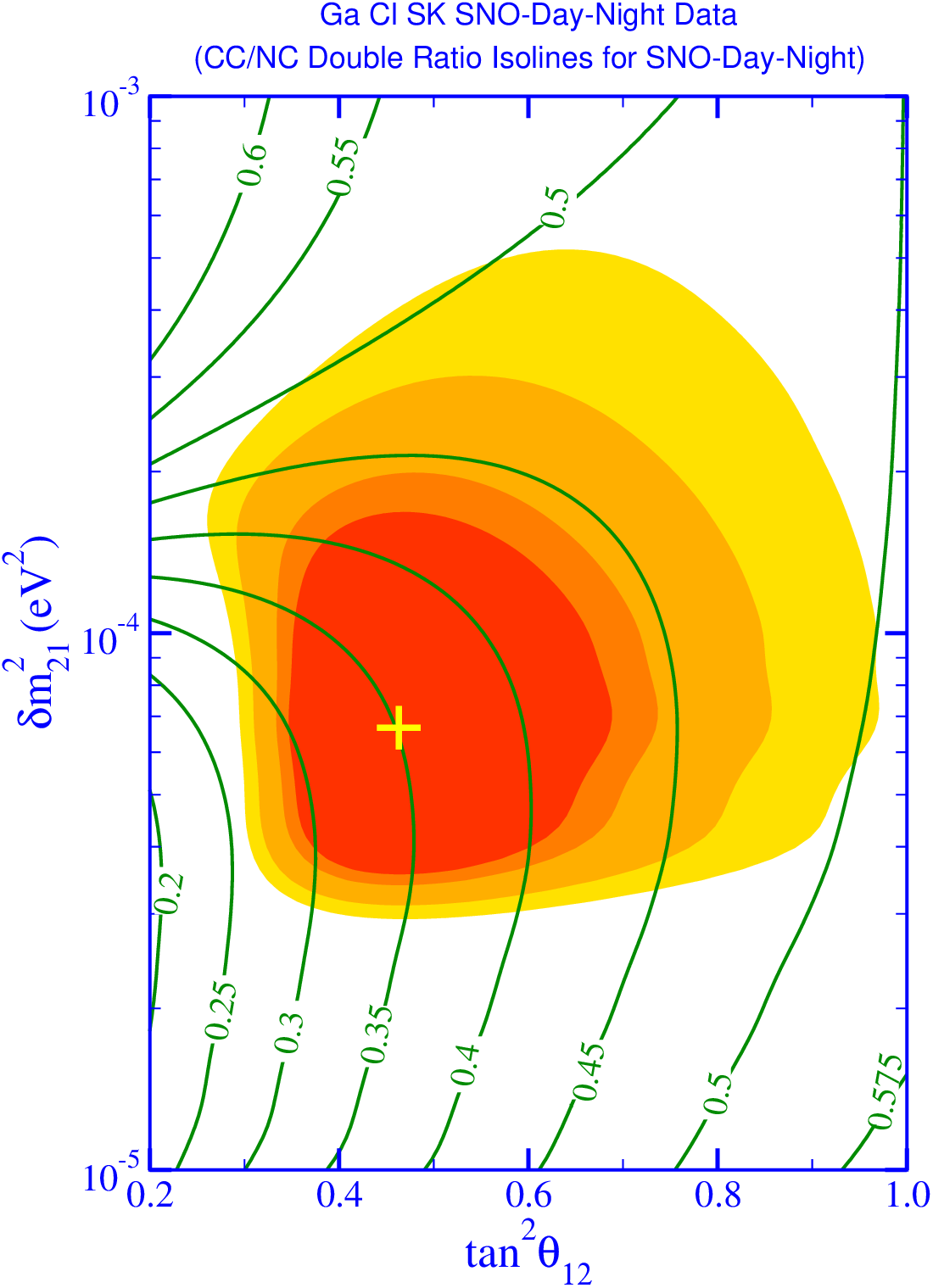}
\vspace*{+2cm} \caption{ \label{fig:1}
Allowed confidence levels in the LMA region of the neutrino 
parameter space when all solar
neutrino experiments (chlorine, average of gallium, 
SNO and SK
experiments) are included except the 
SNO salt phase data.
CC/NC double ratio isolines for SNO experiment 2002 data also 
shown with
$T_{e,th}=5$ MeV.  At best fit (marked by a cross), the value of 
this ratio is $0.35$.
}\end{figure}
\begin{figure}
\includegraphics[scale=.7]{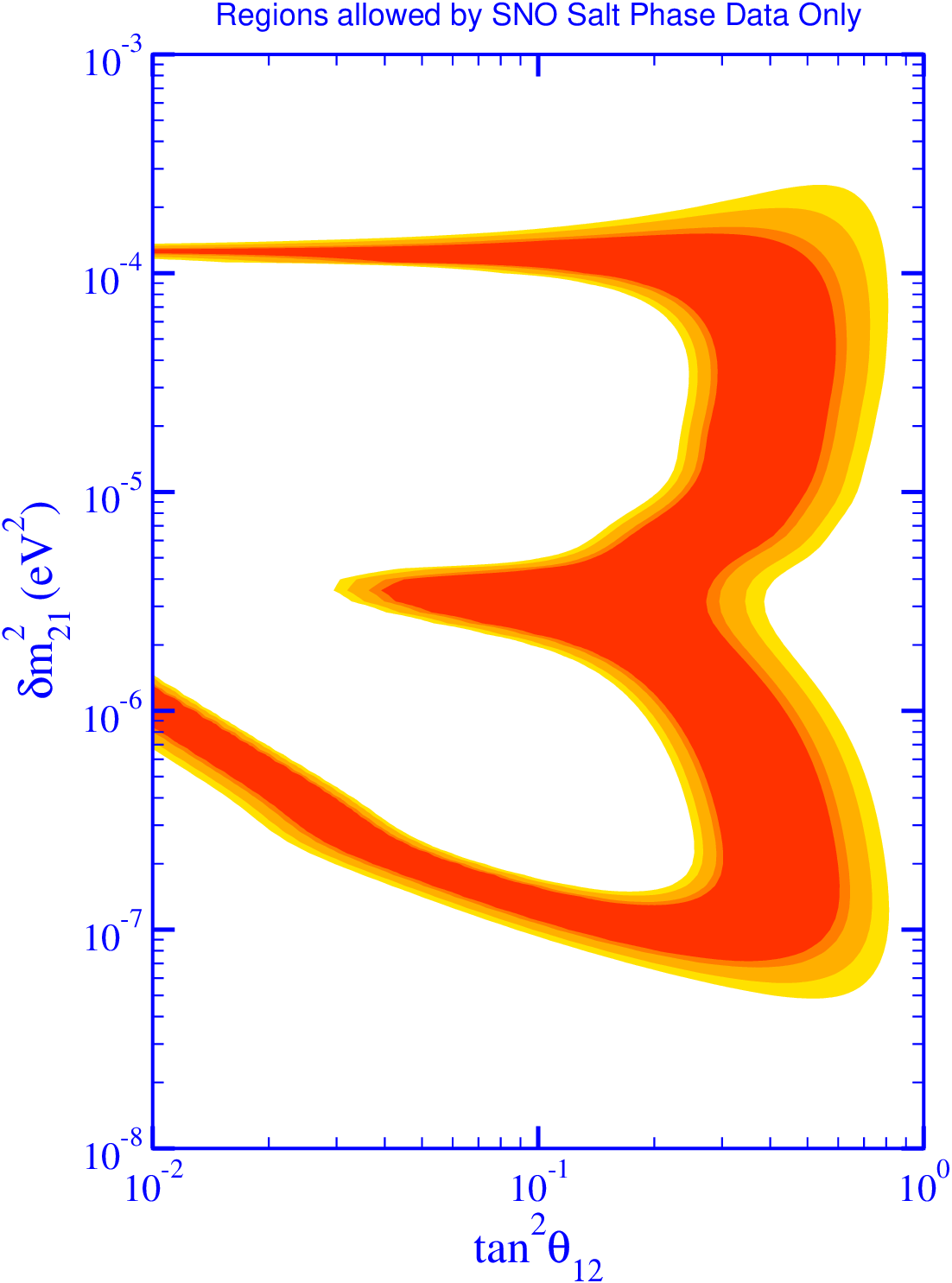}
\vspace*{+2cm} \caption{ \label{fig:2}
Allowed confidence levels in the neutrino parameter 
space from SNO salt phase measurement of NC, CC and ES only.
}\end{figure}
\begin{figure}
\includegraphics[scale=.7]{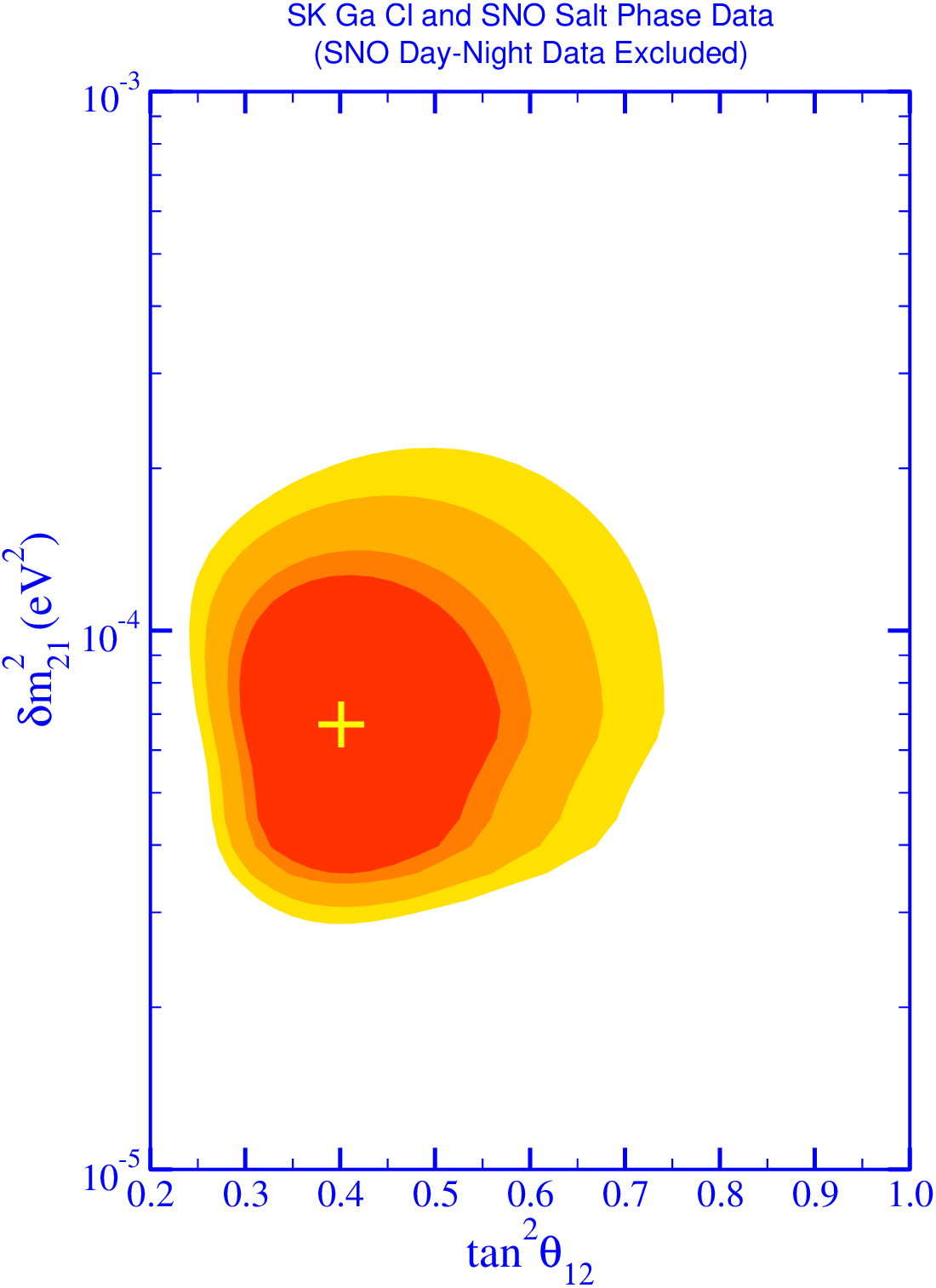}
\vspace*{+2cm} \caption{ \label{fig:3}
Allowed confidence levels in the LMA region of neutrino parameter 
space when chlorine, average rate of gallium experiments, SK 
zenith-spectrum
and  SNO salt phase measurement of NC, CC and ES included in the  
analyses.
SNO day-night spectrum data from 2002 is not included in  order to
better understand the impact of the new salt phase measurement. 
Note that LOW solution is not allowed at 99.73\% C.L. 
in this analysis (not shown). 
The best fit is marked by a cross.
}\end{figure}
\begin{figure}
\includegraphics[scale=.7]{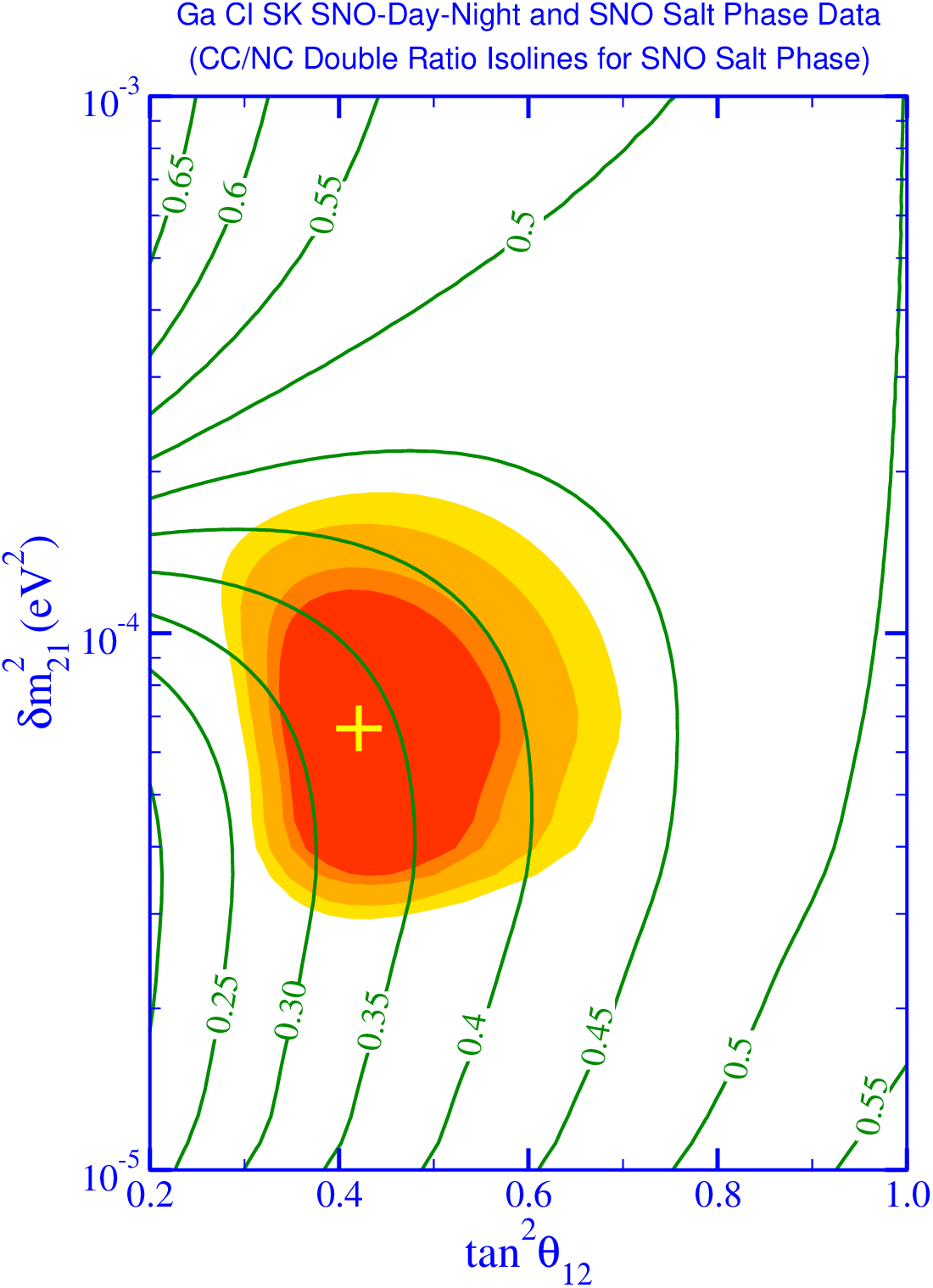}
\vspace*{+2cm} \caption{ \label{fig:4}
Allowed confidence levels of neutrino parameter 
space when all available solar
neutrino data (chlorine, average gallium, SNO and SK
spectrums and  SNO salt phase data) included in the analyses.
CC/NC double ratio isolines for SNO Salt Phase data are 
also shown with
$T_{e,th}=5.5$ MeV. At best fit (marked by a cross), the value of 
this ratio is $0.33$.
}\end{figure}
\begin{figure}
\includegraphics[scale=.7]{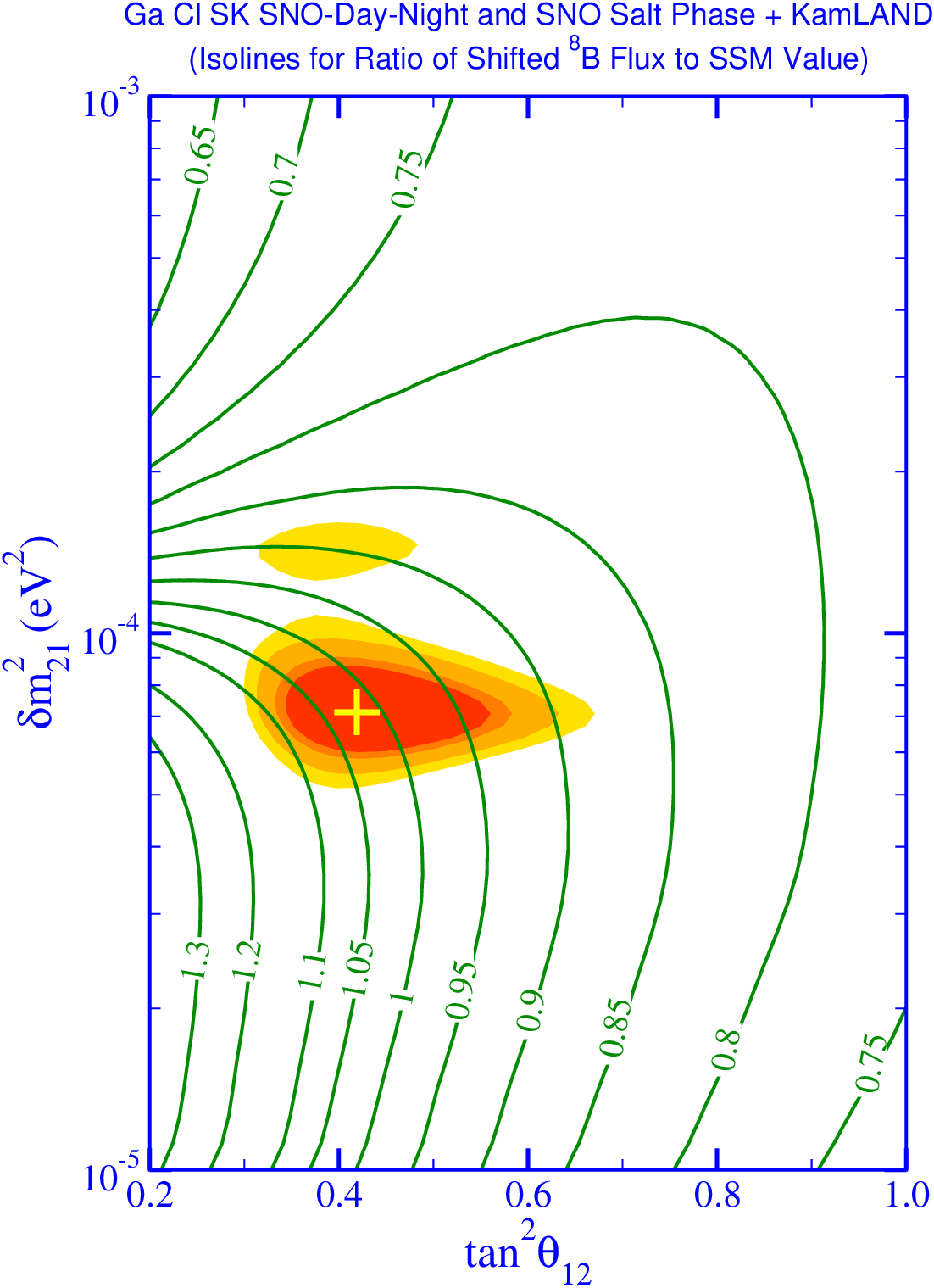}
\vspace*{+2cm} \caption{ \label{fig:6}
Allowed confidence levels from the joint analysis of all
available solar neutrino data (chlorine, average gallium, SNO and SK
spectra and SNO salt phase) and KamLAND reactor data 
The isolines for ratio of the shifted $^8$B flux (as described 
in the text) 
to the SSM value also superimposed on the plot.
At best fit (marked by a cross) the value of this ratio is $1.02$. 
}\end{figure}
\begin{figure}
\includegraphics[scale=.7]{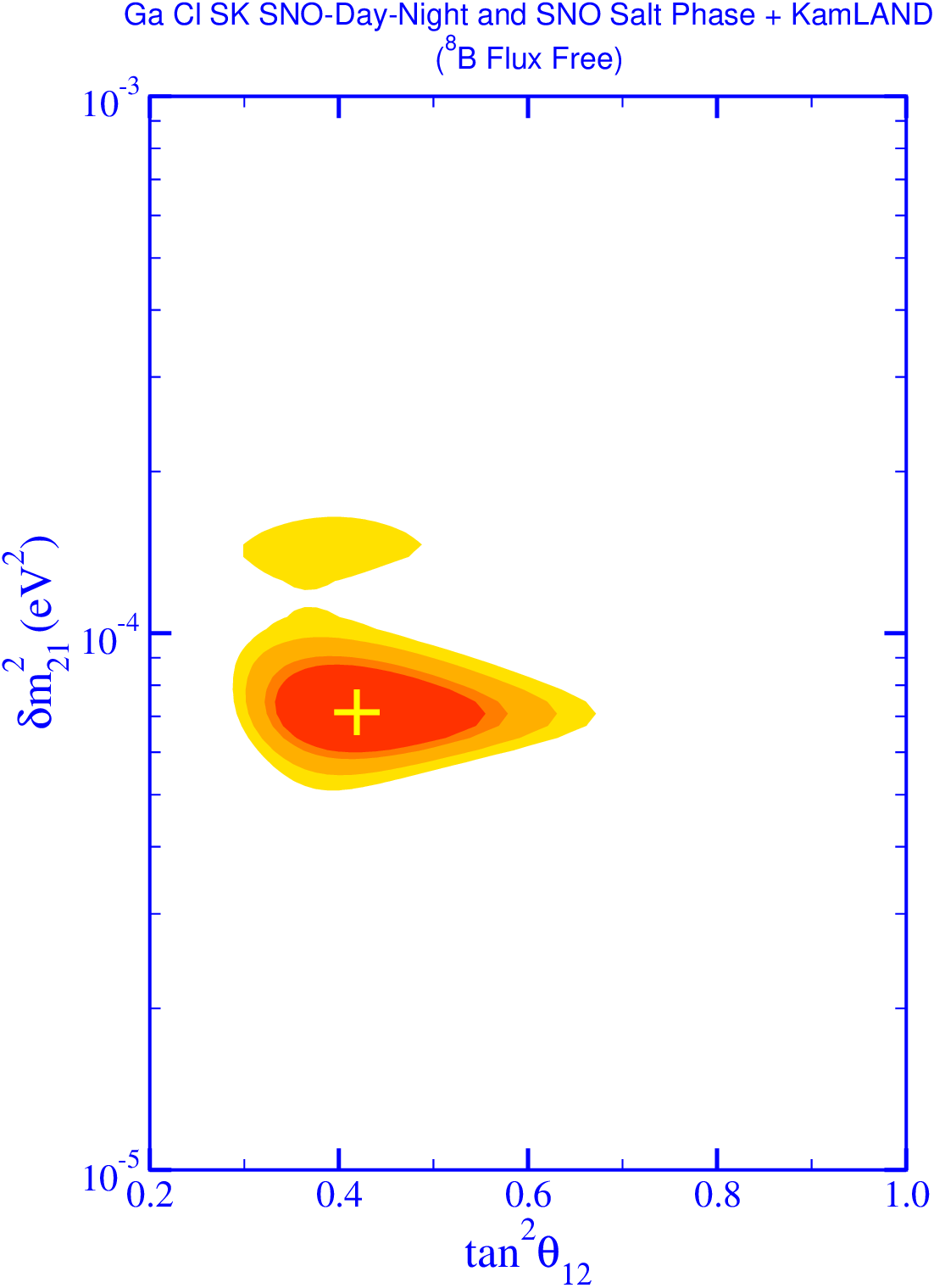}
\vspace*{+2cm} \caption{ \label{fig:7}
Same as the previous figure, except $^8B$ flux is treated as a free 
parameter.
Note that, shape of higher confidence levels look more similar to 
the original
analyses of SNO Collaboration in which $^8B$ is treated as a free 
parameter as well.
}\end{figure}


\begin{thebibliography}{99}


\bibitem{yenisno}
S.~N.~Ahmed {\it et al.} [SNO Collaboration],
arXiv:nucl-ex/0309004.

\bibitem{Junghans:2003bd}
A.~R.~Junghans {\it et al.},
arXiv:nucl-ex/0308003.

\bibitem{Junghans:2001ee}
A.~R.~Junghans {\it et al.},
Phys.\ Rev.\ Lett.\  {\bf 88}, 041101 (2002)
[arXiv:nucl-ex/0111014].

\bibitem{Ahmad:2001an}
Q.~R.~Ahmad {\it et al.}  [SNO Collaboration],
Phys.\ Rev.\ Lett.\  {\bf 87}, 071301 (2001)
[arXiv:nucl-ex/0106015].

\bibitem{Ahmad:2002ka}
Q.~R.~Ahmad {\it et al.}  [SNO Collaboration],
Phys.\ Rev.\ Lett.\  {\bf 89}, 011302 (2002)
[arXiv:nucl-ex/0204009].

\bibitem{Ahmad:2002jz}
Q.~R.~Ahmad {\it et al.}  [SNO Collaboration],
Phys.\ Rev.\ Lett.\  {\bf 89}, 011301 (2002)
[arXiv:nucl-ex/0204008].

\bibitem{Bahcall:2000nu}
J.~N.~Bahcall, M.~H.~Pinsonneault and S.~Basu,
Astrophys.\ J.\  {\bf 555}, 990 (2001)
[arXiv:astro-ph/0010346].

\bibitem{Eguchi:2002dm}
K.~Eguchi {\it et al.}  [KamLAND Collaboration],
Phys.\ Rev.\ Lett.\  {\bf 90}, 021802 (2003)
[arXiv:hep-ex/0212021].


\bibitem{Balantekin:2003dc}
A.~B.~Balantekin and H.~Yuksel,
J.\ Phys.\ G {\bf 29}, 665 (2003)
[arXiv:hep-ph/0301072].

\bibitem{Fogli:2002pt}
G.~L.~Fogli, E.~Lisi, A.~Marrone, D.~Montanino and A.~Palazzo,
Phys.\ Rev.\ D {\bf 66}, 053010 (2002)[arXiv:hep-ph/0206162].

\bibitem{HOWTOSNO}
SNO Collaboration, Data Page, 
http://www.sno.phy.queensu.ca/sno/prlwebpage

\bibitem{HOWTOSNO2}
SNO Collaboration, Data Page, 
http://www.sno.phy.queensu.ca/sno/results-09-03/howto.ps


\bibitem{Abdurashitov:2002nt}
J.~N.~Abdurashitov {\it et al.}  [SAGE Collaboration],
J.\ Exp.\ Theor.\ Phys.\  {\bf 95}, 181 (2002)
[Zh.\ Eksp.\ Teor.\ Fiz.\  {\bf 122}, 211 (2002)]
[arXiv:astro-ph/0204245].

\bibitem{Hampel:1998xg}
W.~Hampel {\it et al.}  [GALLEX Collaboration],
Phys.\ Lett.\ B {\bf 447}, 127 (1999).

\bibitem{Altmann:2000ft}
M.~Altmann {\it et al.}  [GNO Collaboration],
Phys.\ Lett.\ B {\bf 490}, 16 (2000)
[arXiv:hep-ex/0006034].

\bibitem{Cleveland:nv}
B.~T.~Cleveland {\it et al.},
Astrophys.\ J.\  {\bf 496}, 505 (1998).

\bibitem{Fukuda:2002pe}
S.~Fukuda {\it et al.}  [Super-Kamiokande Collaboration],
Phys.\ Lett.\ B {\bf 539}, 179 (2002) 
[arXiv:hep-ex/0205075].

\bibitem{Butler:1999sv}
M.~Butler and J.~W.~Chen,
Nucl.\ Phys.\ A {\bf 675}, 575 (2000)
[arXiv:nucl-th/9905059]; 
M.~Butler, J.~W.~Chen and X.~Kong,
Phys.\ Rev.\ C {\bf 63}, 035501 (2001)
[arXiv:nucl-th/0008032].

\bibitem{Kurylov:2002vj}
A.~Kurylov, M.~J.~Ramsey-Musolf and P.~Vogel,
Phys.\ Rev.\ C {\bf 67}, 035502 (2003)
[arXiv:hep-ph/0211306].

\bibitem{Chen:2002pv}
J.~W.~Chen, K.~M.~Heeger and R.~G.~H. ~Robertson,
Phys.\ Rev.\ C {\bf 67}, 025801 (2003)
[arXiv:nucl-th/0210073].

\bibitem{Balantekin:2003ep}
A.~B.~Balantekin and H.~Yuksel,
arXiv:hep-ph/0307227.

\bibitem{Bahcall:2002zh}
J.~N.~Bahcall, M.~C.~Gonzalez-Garcia and C.~Pena-Garay,
Phys.\ Rev.\ C {\bf 66}, 035802 (2002)
[arXiv:hep-ph/0204194].

\bibitem{Maris:2001tg}
M.~Maris and S.~T.~Petcov,
Phys.\ Lett.\ B {\bf 534}, 17 (2002)
[arXiv:hep-ph/0201087].


\end{thebibliography}
\end{document}